\documentstyle[12pt,epsfig]{article}
\begin{document}

\begin{flushright}
{\bf hep-ph/0204049} \\
{BIHEP-TH-2002-13} 
\end{flushright}

\vspace{0.5cm}

\begin{center}
{\large\bf Nearly Tri-Bimaximal Neutrino Mixing and CP Violation}
\end{center}

\vspace{0.1cm}

\begin{center}
{\bf Zhi-zhong Xing
\footnote{Electronic address: xingzz@mail.ihep.ac.cn}} \\ 
{\it Institute of High Energy Physics, P.O. Box 918, Beijing 100039, China}
\end{center}

\vspace{2.5cm}

\begin{abstract}
We point out two simple but instructive possibilities to modify the
tri-bimaximal neutrino mixing ansatz, such that leptonic CP violation
can naturally be incorporated into the resultant scenarios of
{\it nearly} tri-bimaximal flavor mixing. The consequences of two 
new ans$\rm\ddot{a}$tze on solar, atmospheric and reactor neutrino 
oscillations are analyzed. We also discuss an interesting approach 
to construct lepton mass matrices under permutation symmetry, from 
which one may derive another {\it nearly} tri-bimaximal neutrino mixing 
scenario with no intrinsic CP violation in neutrino oscillations.
\end{abstract}


\newpage

\section{Introduction}

The atmospheric and solar neutrino oscillations observed in the 
Super-Kamiokande experiment \cite{SK} have provided 
robust evidence that neutrinos are massive and lepton flavors are mixed.
All analyses of the atmospheric neutrino deficit favor 
$\nu_\mu \rightarrow \nu_\tau$ as the dominant oscillation 
mode with the mass-squared difference 
$\Delta m^2_{\rm atm} \sim 10^{-3} ~ {\rm eV^2}$ and the
mixing factor $\sin^2 2\theta_{\rm atm} > 0.85$ at the
$99\%$ confidence level. In addition, 
the present Super-Kamiokande and SNO \cite{SNO} data indicate that the
solar neutrino anomaly is most likely attributed to the
matter-enhanced $\nu_e\rightarrow \nu_\mu$ 
oscillation via the Mikheyev-Smirnov-Wolfenstein (MSW) mechanism \cite{MSW}
with $\Delta m^2_{\rm sun} \sim 10^{-5} ~ {\rm eV^2}$ and
$\sin^2 2\theta_{\rm sun} \sim 0.6 - 0.98$ at the
$3\sigma$ confidence level (large-angle MSW solution).
The strong hierarchy between $\Delta m^2_{\rm atm}$ and $\Delta m^2_{\rm sun}$,
together with the small $\nu_3$-component in $\nu_e$ 
configuration restricted by the CHOOZ reactor neutrino
oscillation experiment \cite{CHOOZ},
implies that atmospheric and solar neutrino oscillations decouple 
approximately from each other. Each of them is 
dominated by a single mass scale, which can be set as
$\Delta m^2_{\rm sun} \equiv |m^2_2 - m^2_1|$ or
$\Delta m^2_{\rm atm} \equiv |m^2_3 - m^2_2|$.
The mixing factors of solar, atmospheric and CHOOZ 
neutrino oscillations are simply given by
\begin{eqnarray}
\sin^2 2\theta_{\rm sun} & = & 4 |V_{e1}|^2 |V_{e2}|^2 \; ,
\nonumber \\
\sin^2 2\theta_{\rm atm} & = & 4 |V_{\mu 3}|^2 \left ( 1- 
|V_{\mu 3}|^2 \right ) \; ,
\nonumber \\
\sin^2 2\theta_{\rm chz} & = & 4 |V_{e 3}|^2 \left ( 1- 
|V_{e 3}|^2 \right ) \; ,
\end{eqnarray}
where $V$ is the $3\times 3$ lepton flavor mixing matrix linking
the neutrino mass eigenstates $(\nu_1, \nu_2, \nu_3)$
to the neutrino flavor eigenstates $(\nu_e, \nu_\mu, \nu_\tau)$.
As current experimental data favor 
$\sin^2 2\theta_{\rm chz} \ll \sin^2 2\theta_{\rm sun} \sim 
\sin^2 2\theta_{\rm atm} \sim {\cal O}(1)$, two large flavor mixing 
angles can be drawn from Eq. (1) in a specific parametrization of $V$: 
one between the 2nd and 3rd lepton 
families and the other between the 1st and 2nd lepton families.

So far a number of phenomenological ans$\rm\ddot{a}$tze of lepton flavor 
mixing with two large rotation angles, including the ``democratic''  
ansatz \cite{FX96} and the ``bimaximal'' ansatz \cite{BM}, 
have been proposed and discussed \cite{Review}. In this paper we pay
our particular attention to a new ansatz of the form
(up to a trivial sign or phase rearrangement) 
\begin{equation}
V_0 \; =\; \left ( \matrix{
\frac{\sqrt{2}}{\sqrt{3}}      & \frac{1}{\sqrt{3}}    & 0 \cr
-\frac{1}{\sqrt{6}}    & \frac{1}{\sqrt{3}}   & \frac{1}{\sqrt{2}} \cr
\frac{1}{\sqrt{6}}     & -\frac{1}{\sqrt{3}}  & \frac{1}{\sqrt{2}} \cr} 
\right ) \; ,
\end{equation}
proposed recently by Harrison, Perkins and Scott \cite{Scott}.
This so-called ``tri-bimaximal'' flavor mixing pattern 
predicts $\sin^2 2\theta_{\rm atm} = 1$ and
$\sin^2 2\theta_{\rm sun} = 8/9$, consistent
very well with the atmospheric neutrino oscillation data and the
large-angle MSW solution to the solar neutrino problem. However,
it leads also to $\sin^2 2\theta_{\rm chz} =0$, implying the absence
of both high-energy matter resonances and intrinsic CP violation in
neutrino oscillations. 

The main purpose of this paper is to discuss two simple but 
instructive possibilities to modify the tri-bimaximal neutrino
mixing pattern in Eq. (2), such that CP violation can naturally be 
incorporated into the resultant scenarios of {\it nearly} tri-bimaximal 
flavor mixing. Two specific textures of the charged lepton mass matrix 
are taken into account, in order to obtain small but non-vanishing 
$|V_{e3}|$ or $\sin^2 2\theta_{\rm chz}$. We find that two new 
ans$\rm\ddot{a}$tze have practically indistinguishable consequences on the 
atmospheric neutrino oscillation, but their predictions for 
$\sin^2 2\theta_{\rm sun}$, $\sin^2 2\theta_{\rm chz}$ and
leptonic CP violation are rather different. We also discuss
an interesting approach to construct lepton mass matrices under 
permutation symmetry, from which one may derive another {\it nearly} 
tri-bimaximal neutrino mixing scenario with $|V_{e3}| \neq 0$ but 
with no intrinsic CP violation in neutrino oscillations.

\section{Nearly tri-bimaximal neutrino mixing}

The fact that masses of three active neutrinos are extremely small
is presumably attributed to the Majorana nature of neutrino 
fields \cite{Seesaw}. In this picture, the light (left-handed) neutrino 
mass matrix $M_\nu$ must be symmetric and can be diagonalized by a single
unitary transformation:
\begin{equation}
U^\dagger_\nu M_\nu U^*_\nu \; = \; 
{\rm Diag} \left \{ m_1, m_2, m_3 \right \} \; .
\end{equation}
The charged lepton mass
matrix $M_l$ is in general non-Hermitian, hence the
diagonalization of $M_l$ needs a bi-unitary transformation:
\begin{equation}
U^{\dagger}_l M_l \; \tilde{U}_l \; = \; 
{\rm Diag} \left \{ m_e, m_\mu, m_\tau \right \} \; . 
\end{equation}
The lepton flavor mixing matrix $V$, defined to link the
neutrino mass eigenstates $(\nu_1, \nu_2, \nu_3)$ to the
neutrino flavor eigenstates $(\nu_e, \nu_\mu, \nu_\tau)$,
measures the mismatch between the diagonalization of $M_l$ and 
that of $M_\nu$: $V = U^{\dagger}_l U_\nu$.
Note that $(m_1, m_2, m_3)$ in Eq. (3) and 
$(m_e, m_\mu, m_\tau)$ in Eq. (4) are physical (real and positive) masses
of light neutrinos and charged leptons, respectively.

In the flavor basis where
$M_l$ is diagonal (i.e., $U_l = {\bf 1}$ being a unity matrix),
the flavor mixing matrix is simplified to $V = U_\nu$. The tri-bimaximal
neutrino mixing pattern $U_\nu =V_0$ can then be constructed from the
product of two Euler rotation matrices: 
\begin{eqnarray}
R_{12}(\theta_x) & = & \left ( \matrix{
c_x	& s_x 	& 0 \cr
-s_x	& c_x	& 0 \cr
0	& 0	& 1 \cr} \right ) \; ,
\nonumber \\
R_{23}(\theta_y) & = & \left ( \matrix{
1	& 0	& 0 \cr
0	& c_y	& s_y \cr
0	& -s_y	& c_y \cr} \right ) \; ,
\end{eqnarray}
where $s_x \equiv \sin\theta_x$, $c_y \equiv \cos\theta_y$, and so on.
Taking $\theta_x = \arctan (1/\sqrt{2}) \approx 35.3^\circ$ and
$\theta_y = 45^\circ$, we obtain
\begin{eqnarray}
V_0 & = & R_{23}(\theta_y) \otimes R_{12}(\theta_x) 
\nonumber \\ 
& = & \left ( \matrix{
\frac{\sqrt{2}}{\sqrt{3}}      & \frac{1}{\sqrt{3}}    & 0 \cr
-\frac{1}{\sqrt{6}}    & \frac{1}{\sqrt{3}}   & \frac{1}{\sqrt{2}} \cr
\frac{1}{\sqrt{6}}     & -\frac{1}{\sqrt{3}}  & \frac{1}{\sqrt{2}} \cr} 
\right ) \; .
\end{eqnarray}
The vanishing of the (1,3) element in $V_0$ assures
an exact decoupling between solar ($\nu_e \rightarrow \nu_\mu$) 
and atmospheric ($\nu_\mu \rightarrow \nu_\tau$) neutrino oscillations.
The corresponding neutrino mass matrix $M_\nu$ takes the form
\begin{eqnarray}
M_\nu & = & V_0 \left ( \matrix{
m_1 & 0 & 0 \cr
0 & m_2 & 0 \cr
0 & 0 & m_3 \cr} \right ) V^{\rm T}_0
\nonumber \\ 
& = & \left ( \matrix{
A_\nu - B_\nu -C_\nu  & C_\nu & -C_\nu \cr
C_\nu & A_\nu & B_\nu \cr
-C_\nu & B_\nu & A_\nu \cr} \right ) \; ,
\end{eqnarray}
where 
\begin{eqnarray}
A_\nu & = & \frac{m_3}{2} ~ + ~ \frac{m_1 + 2 m_2}{6}  \;\; , 
\nonumber \\
B_\nu & = & \frac{m_3}{2} ~ - ~ \frac{m_1 + 2 m_2}{6}  \;\; , 
\nonumber \\
C_\nu & = & \frac{m_2 - m_1}{3} \;\; .
\end{eqnarray}
If $m_1 \approx m_2$ holds, one may arrive at 
a simpler texture of $M_\nu$ with $A_\nu \approx (m_3 + m_1)/2$, 
$B_\nu \approx (m_3 - m_1)/2$ and $C_\nu \approx 0$.

Let us comment briefly on the mathematical structure of $M_\nu$
obtained in Eq. (7). Indeed $M_\nu$ can be decomposed as
\begin{equation}
M_\nu \; =\; A_\nu {\bf I}_A + B_\nu {\bf I}_B + C_\nu {\bf I}_C \; ,
\end{equation}
where
\begin{eqnarray}
{\bf I}_A & = & \left ( \matrix{
~ 1 ~	& ~ 0 ~	& ~ 0 \cr
~ 0 ~	& ~ 1 ~	& ~ 0 \cr
~ 0 ~	& ~ 0 ~	& ~ 1 \cr} \right ) \; ,
\nonumber \\
{\bf I}_B & = & \left ( \matrix{
-1	& 0 ~	& ~ 0 \cr
0	& 0 ~	& ~ 1 \cr
0	& 1 ~	& ~ 0 \cr} \right ) \; ,
\nonumber \\
{\bf I}_C & = & \left ( \matrix{
-1	& 1	& -1 \cr
1	& 0	& 0 \cr
-1	& 0	& 0 \cr} \right ) \; .
\end{eqnarray}
Such a structure of the neutrino mass matrix is very similar to
that giving rise to the bimaximal flavor mixing \cite{Xing01}. 
Note that the diagonalization of $M_\nu$ requires an
orthogonal matrix which is able to diagonalize ${\bf I}_B$ and
${\bf I}_C$ simultaneously. This orthogonal matrix is just
$V_0$ given in Eq. (6). Although the decomposition of $M_\nu$
shown above is by no means unique, it might have a meaningful
interpretation in an underlying theory of neutrino masses with
specific flavor symmetries.

The tri-bimaximal neutrino mixing pattern will be
modified, if $U_l$ deviates somehow from the unity matrix. This
can certainly happen, provided that the charged lepton mass matrix $M_l$ is
not diagonal in the flavor basis where the neutrino mass matrix
$M_\nu$ takes the form given in Eq. (7). 
As $U_\nu = V_0$ describes a product of 
two special Euler rotations in the real (2,3) and
(1,2) planes, the simplest form of $U_l$ which allows $V=U^\dagger_l U_\nu$ 
to cover the whole $3\times 3$ space should be 
$U_l = R_{12}(\theta_x)$ or $U_l = R_{31}(\theta_z)$ 
(see Ref. \cite{Xing98} for a detailed discussion). 
To make CP violation incorporated into $V$, we adopt 
the complex rotation matrices: 
\begin{eqnarray}
R_{12}(\theta_x, \phi_x) & = & \left ( \matrix{
c_x	& s_x e^{i\phi_x}	& 0 \cr
-s_x e^{-i\phi_x}	& c_x	& 0 \cr
0	& 0	& 1 \cr} \right ) \; ,
\nonumber \\
R_{31}(\theta_z, \phi_z) & = & \left ( \matrix{
c_z	& 0	& s_z e^{i\phi_z} \cr
0	& 1	& 0 \cr
-s_z e^{-i\phi_z}	& 0	& c_z	\cr} \right ) \; .
\end{eqnarray}
In this case, we arrive at
lepton flavor mixing of the pattern 
\begin{eqnarray}
V_{(x)} & = & R^\dagger_{12}(\theta_x, \phi_x) \otimes V_0 
\nonumber \\
& = & \left ( \matrix{
\frac{1}{\sqrt{6}} \left ( 2 c_x + s_x e^{i\phi_x} \right ) & 
\frac{1}{\sqrt{3}} \left ( c_x - s_x e^{i\phi_x} \right) & 
-\frac{1}{\sqrt{2}} s_x e^{i\phi_x} \cr
-\frac{1}{\sqrt{6}} \left ( c_x - 2 s_x e^{-i\phi_x} \right ) & 
\frac{1}{\sqrt{3}} \left ( c_x + s_x e^{-i\phi_x} \right ) &
\frac{1}{\sqrt{2}} c_x \cr
\frac{1}{\sqrt{6}}  & -\frac{1}{\sqrt{3}}   & \frac{1}{\sqrt{2}} \cr} 
\right ) \; ,
\end{eqnarray}
or of the pattern
\begin{eqnarray}
V_{(z)} & = & R^\dagger_{31}(\theta_z, \phi_z) \otimes V_0
\nonumber \\ 
& = & \left ( \matrix{ 
\frac{1}{\sqrt{6}} \left ( 2 c_z - s_z e^{i\phi_z} \right ) & 
\frac{1}{\sqrt{3}} \left ( c_z + s_z e^{i\phi_z} \right ) & 
-\frac{1}{\sqrt{2}} s_z e^{i\phi_z} \cr
-\frac{1}{\sqrt{6}}    & \frac{1}{\sqrt{3}}    & \frac{1}{\sqrt{2}} \cr
\frac{1}{\sqrt{6}} \left ( c_z + 2 s_z e^{-i\phi_z} \right ) & 
-\frac{1}{\sqrt{3}} \left ( c_z - s_z e^{-i\phi_z} \right ) &
\frac{1}{\sqrt{2}} c_z \cr} \right ) \; .
\end{eqnarray}
It is obvious that $V_{(x)}$ and $V_{(z)}$ represent two 
nearly tri-bimaximal flavor mixing scenarios, if the rotation 
angles $\theta_x$ and $\theta_z$ are small in magnitude. The
complex phase $\phi_x$ in $V_{(x)}$ or $\phi_z$ in $V_{(z)}$ is 
the source of leptonic CP violation in neutrino oscillations.

\section{Constraints on mixing factors and CP violation}

As the mixing angle $\theta_x$ or $\theta_z$ arises from
the diagonalization of $M_l$, it is expected to be a simple function
of the ratios of charged lepton masses. Then the strong mass
hierarchy of charged leptons naturally assures the
smallness of $\theta_x$ or $\theta_z$, as one can see later on.

Indeed a proper texture of $M_l$ 
which may lead to the flavor mixing pattern $V_{(x)}$ is 
\begin{equation}
M^{(x)}_l \; =\; \left ( \matrix{
0  & C_l & 0 \cr
C^*_l & B_l  & 0 \cr
0 & 0 & A_l \cr} \right ) \;\; ,
\end{equation}
where $A_l = m_\tau$, $B_l = m_\mu - m_e$, and 
$C_l =\sqrt{m_e m_\mu} ~ e^{i\phi_x}$.
The mixing angle $\theta_x$ in $V_{(x)}$ is then given by
\begin{equation}
\tan (2\theta_x) \;\; =\;\; 2 ~ \frac{\sqrt{m_e m_\mu}}{m_\mu - m_e} \;\; .
\end{equation}
On the other hand, a proper texture of $M_l$ which may give rise to
the mixing pattern $V_{(z)}$ reads as follows: 
\begin{equation}
M^{(z)}_l \; =\; \left ( \matrix{
0  & 0 & C_l \cr
0  & B_l  & 0 \cr
C^*_l & 0 & A_l \cr} \right ) \;\; ,
\end{equation}
where $A_l = m_\tau -m_e$, $B_l = m_\mu$, and 
$C_l =\sqrt{m_e m_\tau} ~ e^{i\phi_z}$.
The mixing angle $\theta_z$ in $V_{(z)}$ turns out to be
\begin{equation}
\tan (2\theta_z) \;\; =\;\; 2 ~ \frac{\sqrt{m_e m_\tau}}{m_\tau - m_e} \;\; .
\end{equation}
In view of the hierarchy of three charged lepton masses 
(i.e., $m_e \ll m_\mu \ll m_\tau$), we obtain 
$s_x \approx \sqrt{m_e/m_\mu}$ and $s_z \approx \sqrt{m_e/m_\tau}$ 
to a good degree of accuracy. Numerically, we find 
$\theta_x \approx 3.978^\circ$ and
$\theta_z \approx 0.972^\circ$ by using the inputs 
$m_e = 0.511$ MeV, $m_\mu = 105.658$ MeV, and $m_\tau = 1.777$ GeV \cite{PDG}. 

Now let us calculate the mixing factors of solar,
atmospheric and reactor neutrino oscillations.
With the help of Eqs. (1) and (12) or (13), we arrive straightforwardly at
\begin{eqnarray}
\sin^2 2\theta^{(x)}_{\rm sun} & = & \frac{8}{9} \left ( 1 - \frac{3}{4} s^2_x
- s_x c_x \cos\phi_x + \frac{3}{2} s^3_x c_x \cos\phi_x 
- 2 s^2_x c^2_x \cos^2\phi_x \right ) \; ,
\nonumber \\
\sin^2 2\theta^{(x)}_{\rm atm} & = & 1 - s^4_x \; ,
\nonumber \\
\sin^2 2\theta^{(x)}_{\rm chz} & = & 1 - c^4_x \; ;
\end{eqnarray}
and 
\begin{eqnarray}
\sin^2 2\theta^{(z)}_{\rm sun} & = & \frac{8}{9} \left ( 1 - \frac{3}{4} s^2_z
+ s_z c_z \cos\phi_z - \frac{3}{2} s^3_z c_z \cos\phi_z 
- 2 s^2_z c^2_z \cos^2\phi_z \right ) \; ,
\nonumber \\
\sin^2 2\theta^{(z)}_{\rm atm} & = & 1 \; ,
\nonumber \\
\sin^2 2\theta^{(z)}_{\rm chz} & = & 1 - c^4_z \; .
\end{eqnarray}
Allowing $\phi_x$ and $\phi_z$ to take 
arbitrary values, we find that the minimal and maximal magnitudes of
$\sin^2 2\theta^{(x)}_{\rm sun}$ and $\sin^2 2\theta^{(z)}_{\rm sun}$
are 
\begin{eqnarray}
\left [\sin^2 2\theta^{(x)}_{\rm sun} \right ]_{\rm min} & = &
\frac{8}{9} \left ( 1 - \frac{3}{4} s^2_x
- s_x c_x + \frac{3}{2} s^3_x c_x - 2 s^2_x c^2_x \right ) \; , 
\nonumber \\
\left [\sin^2 2\theta^{(x)}_{\rm sun} \right ]_{\rm max} & = &
\frac{8}{9} \left ( 1 - \frac{3}{4} s^2_x
+ s_x c_x - \frac{3}{2} s^3_x c_x - 2 s^2_x c^2_x \right ) \; ;
\end{eqnarray}
and
\begin{eqnarray}
\left [\sin^2 2\theta^{(z)}_{\rm sun} \right ]_{\rm min} & = &
\frac{8}{9} \left ( 1 - \frac{3}{4} s^2_z
- s_z c_z + \frac{3}{2} s^3_z c_z - 2 s^2_z c^2_z \right ) \; , 
\nonumber \\
\left [\sin^2 2\theta^{(z)}_{\rm sun} \right ]_{\rm max} & = &
\frac{8}{9} \left ( 1 - \frac{3}{4} s^2_z
+ s_z c_z - \frac{3}{2} s^3_z c_z - 2 s^2_z c^2_z \right ) \; ,
\end{eqnarray}
respectively. A numerical illustration of $\sin^2 2\theta^{(x)}_{\rm sun}$
and $\sin^2 2\theta^{(z)}_{\rm sun}$ as functions of $\phi_x$ and $\phi_z$
is shown in Fig. 1, from which 
$0.816 \leq \sin^2 2\theta^{(x)}_{\rm sun} \leq 0.938$ and
$0.873 \leq \sin^2 2\theta^{(z)}_{\rm sun} \leq 0.903$ can be obtained.
Note that $\sin^2 2\theta_{\rm atm} = 1.000$ holds in both scenarios. 
In addition, we get 
$\sin^2 2\theta^{(x)}_{\rm chz} \approx 0.01$ and
$\sin^2 2\theta^{(z)}_{\rm chz} \approx 0.0006$. 
Therefore two nearly tri-bimaximal neutrino mixing patterns
are practically indistinguishable in the atmospheric neutrino oscillation
experiment. It is possible to distinguish 
between them in the solar neutrino oscillation experiment. 
They can unambiguously be distinguished with the measurements 
of $|V_{e3}|$ and CP or T violation in a variety of long-baseline
neutrino oscillation experiments. 

The strength of CP or T violation in neutrino oscillations,
no matter whether neutrinos are Dirac or Majorana particles,
is measured by a universal and rephasing-invariant parameter
$\cal J$ \cite{Jarlskog}, defined through the following equation:
\begin{equation}
{\rm Im} \left (V_{\alpha i} V_{\beta j} V^*_{\alpha j} V^*_{\beta i} \right ) 
\; =\; {\cal J} \sum_{\gamma, k} \left (\varepsilon_{\alpha\beta\gamma}
\varepsilon_{ijk} \right ) \; ,
\end{equation}
in which the Greek subscripts run over $(e, \mu, \tau)$, and the 
Latin subscripts run over $(1,2,3)$.
Considering two lepton mixing scenarios proposed above, we obtain 
\begin{eqnarray}
{\cal J}_{(x)} & = & \frac{1}{6} s_x c_x \sin \phi_x \; ,
\nonumber \\
{\cal J}_{(z)} & = & \frac{1}{6} s_z c_z \sin \phi_z \; .
\end{eqnarray}
For illustration,
we typically take $\phi_x = \phi_z = 90^\circ$. Then
we arrive at ${\cal J}_{(x)} \approx 0.0115$ and
${\cal J}_{(z)} \approx 0.0028$, respectively.
The former could be determined from the CP-violating asymmetry 
between $\nu_\mu \rightarrow \nu_e$ and 
$\bar{\nu}_\mu \rightarrow \bar{\nu}_e$ transitions or from
the T-violating asymmetry between $\nu_\mu \rightarrow \nu_e$ and 
$\nu_e \rightarrow \nu_\mu$ transitions in a long-baseline neutrino 
oscillation experiment, if the terrestrial matter effects are 
insignificant or under control.

\section{Further discussions and remarks}

We have discussed two simple possibilities to construct the charged
lepton and neutrino mass matrices, from which two nearly tri-bimaximal
neutrino mixing patterns can naturally emerge. Both scenarios
are compatible with the large-angle MSW solution to the solar neutrino 
problem, although their numerical predictions for the mixing factor 
$\sin^2 2\theta_{\rm sun}$ may be different from each other. 
Two lepton mixing patterns are practically indistinguishable in the
atmospheric neutrino oscillation experiment, but their consequences
on $|V_{e3}|$ and leptonic CP violation are different and distinguishable. 
Only one of them is likely to yield an observable CP- or T-violating 
asymmetry in long-baseline neutrino oscillation experiments. 

There are certainly other possibilities to modify the 
tri-bimaximal neutrino mixing ansatz, such that non-vanishing 
$|V_{e3}|$ (and CP violation) can naturally be incorporated into  
the resultant scenarios of nearly tri-bimaximal mixing. For
illustration, we follow an interesting approach proposed in 
Ref. \cite{Scott} to consider charged lepton and neutrino mass 
matrices of the form
\begin{eqnarray}
M_l M^\dagger_l & = & \left ( \matrix{
a	& b	& b^* \cr
b^*	& a	& b \cr
b	& b^*	& a \cr} \right ) \; ,
\nonumber \\
M_\nu M^\dagger_\nu & = & \left ( \matrix{
x	& 0	& ~ y \cr
0	& z	& ~ 0 \cr
y^*	& 0	& ~ x \cr} \right ) \; .
\end{eqnarray}
Clearly $M_l M^\dagger_l$ is invariant under cyclic permutation of
three generation indices \cite{P}, and $M_\nu M^\dagger_\nu$ has
four texture zeros \cite{Zero}.
Note that $y$ was assumed to be real in Ref. \cite{Scott}.
One will see later on that $|V_{e3}| \neq 0$ may non-trivially result from
the phase of $y$. The Hermitian matrices $M_l M^\dagger_l$ 
and $M_\nu M^\dagger_\nu$ can be diagonalized as follows:
\begin{eqnarray}
U^\dagger_l M_l M^\dagger_l U_l & = & {\rm Diag} \left \{ m^2_e,
m^2_\mu, m^2_\tau \right \} \; ,
\nonumber \\
U^\dagger_\nu M_\nu M^\dagger_\nu U_\nu & = & {\rm Diag}
\left \{ m^2_1, m^2_2, m^2_3 \right \} \; .
\end{eqnarray}
It is straightfoward to obtain
\begin{equation}
U_l \; = \; \left ( \matrix{
\frac{1}{\sqrt{3}}	
& \frac{\bar{\omega}}{\sqrt{3}}	
& \frac{\omega}{\sqrt{3}} \cr
\frac{1}{\sqrt{3}}	
& \frac{1}{\sqrt{3}}	
& \frac{1}{\sqrt{3}} \cr
\frac{1}{\sqrt{3}}	
& \frac{\omega}{\sqrt{3}}	
& \frac{\bar{\omega}}{\sqrt{3}} \cr} \right ) \; ,
\end{equation}
where $\omega = \exp (+i 2\pi/3)$ and $\bar{\omega} = \exp (-i 2\pi/3)$;
and 
\begin{equation}
U_\nu \; = \; \left ( \matrix{
\frac{1}{\sqrt{2}} 	& 0	& -\frac{1}{\sqrt{2}} e^{i\varphi} \cr
0	& 1	& 0 \cr
\frac{1}{\sqrt{2}} e^{-i\varphi}	& 0	& \frac{1}{\sqrt{2}} \cr}
\right ) \; ,
\end{equation}
where $\varphi = \arg (y)$.
Then the lepton flavor mixing matrix $V = U^\dagger_l U_\nu$, which describes
the mismatch between the diagonalization of $M_l M^\dagger_l$ and that
of $M_\nu M^\dagger_\nu$, takes the form
\begin{equation}
V \; =\; \left ( \matrix{
\frac{1}{\sqrt{6}} \left ( 1 + e^{-i\varphi} \right )
& \frac{1}{\sqrt{3}} 
& \frac{1}{\sqrt{6}} \left ( 1 - e^{i\varphi} \right ) \cr
\frac{1}{\sqrt{6}} \left ( \omega + \bar{\omega} e^{-i\varphi} \right )
& \frac{1}{\sqrt{3}} 
& \frac{1}{\sqrt{6}} \left ( \bar{\omega} - \omega e^{i\varphi} \right ) \cr
\frac{1}{\sqrt{6}} \left ( \bar{\omega} + \omega e^{-i\varphi} \right )
& \frac{1}{\sqrt{3}} 
& \frac{1}{\sqrt{6}} \left ( \omega - \bar{\omega} e^{i\varphi} \right ) \cr}
\right ) \; .
\end{equation}
The tri-bimaximal neutrino mixing (up to a trivial sign or
phase rearrangement \cite{Scott}) can then be reproduced from $V$ in
the limit $\varphi = 0$. The nearly tri-bimaximal mixing scenario 
obtained in Eq. (28) leads to
\begin{eqnarray}
\sin^2 2\theta_{\rm sun} & = & \frac{8}{9} \cos^2\frac{\varphi}{2} \;\; ,
\nonumber \\
\sin^2 2\theta_{\rm atm} & = & \frac{8}{3} \sin^2 \left ( \frac{\varphi}{2} 
+ \frac{2\pi}{3} \right ) \left [ 1 - \frac{2}{3} \sin^2
\left ( \frac{\varphi}{2} + \frac{2\pi}{3} \right ) \right ] \; ,
\nonumber \\
\sin^2 2\theta_{\rm chz} & = & \frac{8}{3} \sin^2 \frac{\varphi}{2} 
\left ( 1 - \frac{2}{3} \sin^2 \frac{\varphi}{2} \right ) \; .
\end{eqnarray}
We plot the changes of $\sin^2 2\theta_{\rm sun}$, 
$\sin^2 2\theta_{\rm atm}$ and $\sin^2 2\theta_{\rm chz}$ as functions 
of $\varphi$ in Fig. 2, where the experimental upper bound
$\sin^2 2 \theta_{\rm chz} < 0.1$ has been taken into account.
One can see that the allowed ranges of $\varphi$ are
$0 \leq \varphi \leq 0.125\pi$ (or $22.5^\circ$) and
$2\pi \geq \varphi \geq 1.875\pi$ (or $337.5^\circ$). Accordingly,
we obtain $0.939 \leq \sin^2 2\theta_{\rm atm} \leq 1$ and
$0.962 \leq \sin^2 2\theta_{\rm atm} \leq 1$. We also obtain
$0.855 \leq \sin^2 2\theta_{\rm sun} \leq 0.889$ for both ranges
of $\varphi$. 

Note that ${\cal J} = 0$ holds exactly, although $V$ is complex.
Therefore no intrinsic CP violation could be observed in neutrino
oscillation experiments, if the tri-bimaximal
neutrino mixing pattern or its revised version in Eq. (28)
were correct. The result ${\cal J} =0$ makes such a nearly tri-bimaximal 
ansatz less interesting. Of course, the complex phases in $V$ may
have significant effects on the neutrinoless double beta decay, if
neutrinos are Majorana particles. 

Finally let us remark that both the tri-bimaximal mixing
pattern and its possible extensions require some peculiar flavor
symmetries to be imposed on the charged lepton and neutrino mass
matrices. It is likely that one of the three nearly tri-bimaximal
neutrino mixing patterns under discussion serves as the
leading-order approximation of a more complicated flavor mixing
matrix. For the time being, however, such simple ans$\rm\ddot{a}$tze
are very instructive and useful to explore the main features of
lepton flavor mixing and CP violation through neutrino oscillations.
We expect that more delicate neutrino oscillation experiments in
the near future can help pin down the explicit pattern of 
neutrino mixing, from which one may get some insight into the
underlying flavor symmetry and its breaking mechanism responsible 
for the origin of both lepton masses and leptonic CP violation.

\vspace{0.5cm}

The author would like to thank H. Fritzsch for useful discussions and
comments. This work was supported in part by the National Natural
Science Foundation of China.

\newpage

\begin{figure}
\vspace{-2.5cm}
\epsfig{file=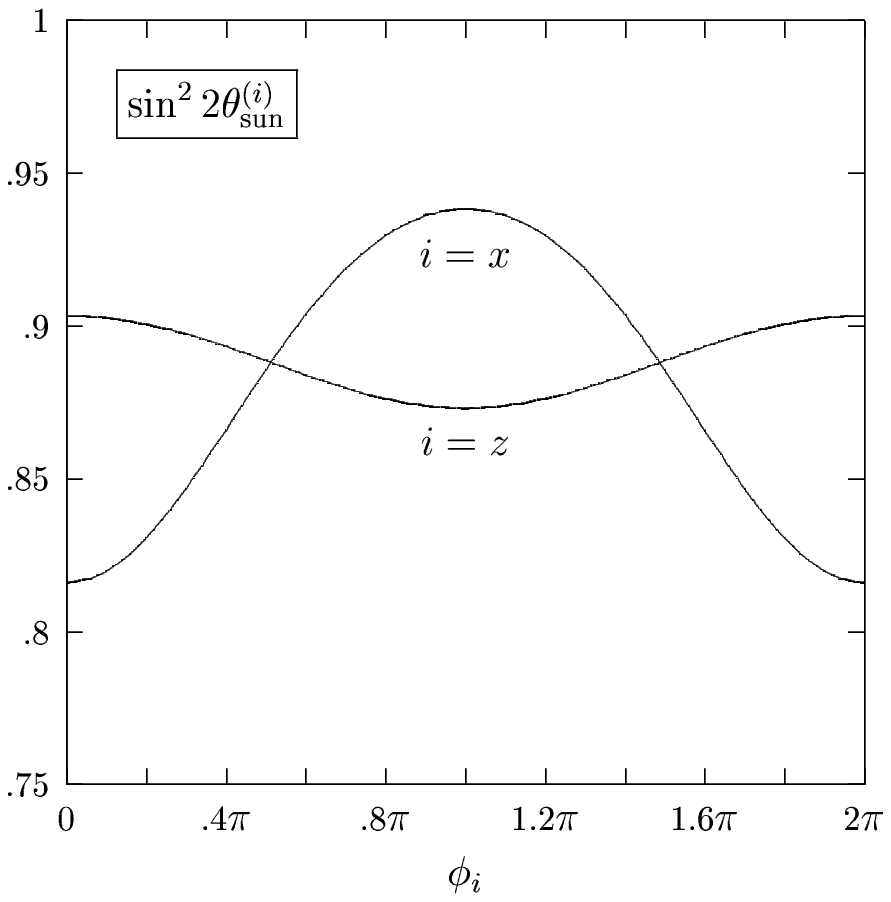,bbllx=-1cm,bblly=4cm,bburx=20cm,bbury=32cm,%
width=16cm,height=22cm,angle=0,clip=}
\vspace{-11.2cm}
\caption{The mixing factors $\sin^2 2\theta^{(x)}_{\rm sun}$ and
$\sin^2 2\theta^{(z)}_{\rm sun}$ against arbitrary values of
$\phi_x$ and $\phi_z$ in two nearly tri-bimaximal neutrino mixing patterns.}
\end{figure}

\begin{figure}
\vspace{-3.5cm}
\epsfig{file=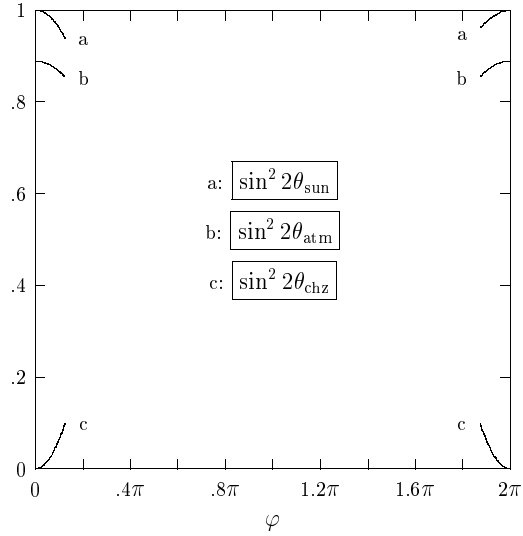,bbllx=-1cm,bblly=4cm,bburx=20cm,bbury=32cm,%
width=16cm,height=22cm,angle=0,clip=}
\vspace{-11.2cm}
\caption{The mixing factors $\sin^2 2\theta_{\rm sun}$,
$\sin^2 2\theta_{\rm atm}$ and $\sin^2 2\theta_{\rm chz}$
against arbitrary values of $\varphi$ in a nearly tri-bimaximal
neutrino mixing scenario.}
\end{figure}


\newpage

\begin{thebibliography}{99}
\bibitem{SK} Y. Fukuda {\it et al.}, Phys. Lett. B {\bf 436} (1998) 33;
Phys. Rev. Lett. {\bf 81} (1998) 1158; {\bf 81} (1998) 1562;
{\bf 82} (1999) 1810; {\bf 85} (2000) 3999.

\bibitem{SNO} SNO Collaboration, Q.R. Ahmad {\it et al.}, 
Phys. Rev. Lett. {\bf 87} (2001) 071301.

\bibitem{MSW} L. Wolfenstein, Phys. Rev. D {\bf 17} (1978) 2369;
S.P. Mikheyev and A.Yu. Smirnov, Sov. J. Nucl. Phys. {\bf 42} (1985) 913.

\bibitem{CHOOZ} CHOOZ Collaboration, 
M. Apollonio {\it et al.}, Phys. Lett. B {\bf 420} (1998) 397; 
Palo Verde Collaboration, F. Boehm {\it et al.}, 
Phys. Rev. Lett. {\bf 84} (2000) 3764.

\bibitem{FX96} H. Fritzsch and Z.Z. Xing,
Phys. Lett. B {\bf 372} (1996) 265;
Phys. Lett. B {\bf 440} (1998) 313;
Phys. Rev. D {\bf 61} (2000) 073016.

\bibitem{BM} F. Vissani, hep-ph/9708843 (unpublished); 
V. Barger, S. Pakvasa, T.J. Weiler, and
K. Whisnant, Phys. Lett. B {\bf 437} (1998) 107; 
D.V. Ahluwalia, Mod. Phys. Lett. A {\bf 13} (1998) 2249;
H. Fritzsch and Z.Z. Xing, Phys. Lett. B {\bf 440} (1998) 313;
A. Baltz, A.S. Goldhaber, and M. Goldhaber, 
Phys. Rev. Lett. {\bf 81} (1998) 5730;
T. Kitabayashi and M. Yasue, Nucl. Phys. B {\bf 609} (2001) 61;
K.S. Babu and R.N. Mohapatra, hep-ph/0201176;
and references therein.

\bibitem{Review} For a review of various neutrino mixing 
ans$\rm\ddot{a}$tze, see:
G. Altarelli and F. Feruglio, Phys. Rep. {\bf 320} (1999) 295; 
H. Fritzsch and Z.Z. Xing, Prog. Part. Nucl. Phys. {\bf 45} (2000) 1; 
S.M. Barr and I. Dorsner, Nucl. Phys. B {\bf 585} (2000) 79.

\bibitem{Scott} P.F. Harrison, D.H. Perkins, and W.G. Scott,
hep-ph/0202074 (to be published in Phys. Lett. B). See also 
W.G. Scott, Nucl. Phys. B (Proc. Suppl.) {\bf 85} (2000) 177;
P.F. Harrison and W.G. Scott, hep-ph/0203209.

\bibitem{Seesaw} T. Yanagida, in {\it Proceedings of the Workshop on Unified 
Theory and Baryon Number in the Universe}, edited by
O. Sawada and A. Sugamoto, KEK, Japan, 1979; 
M. Gell-Mann, P. Ramond, and R. Slansky,
in {\it Supergravity}, edited by D. Freedman and P. van Nieuwenhuizen
(North Holland, Amsterdam, 1979), p. 315; 
R. Mohapatra and G. Senjanovic, Phys. Rev. Lett. {\bf 44} (1980) 912.

\bibitem{Xing01} Z.Z. Xing, Phys. Rev. D {\bf 64} (2001) 093013.

\bibitem{Xing98} C. Jarlskog, in {\it CP Violation}, edited by
C. Jarlskog (World Scientific, Singapore, 1989), p. 3;
H. Fritzsch and Z.Z. Xing, Phys. Rev. D {\bf 57} (1998) 574.

\bibitem{PDG} Particle Data Group, D.E. Groom {\it et al.},
Eur. Phys. J. C {\bf 15} (2000) 1.

\bibitem{Jarlskog} C. Jarlskog, Phys. Rev. Lett. {\bf 55} (1985) 1039.

\bibitem{P} See, e.g., 
P.F. Harrison and W.G. Scott, Phys. Lett. B {\bf 333} (1994) 471; 
S.L. Adler, Phys. Rev. D {\bf 59} (1999) 015012;

\bibitem{Zero} See, e.g., H. Fritzsch, Phys. Lett. B {\bf 73} (1978) 317;
Nucl. Phys. B {\bf 155} (1979) 189;
P. Ramond, R.G. Roberts, and G.G. Ross, Nucl. Phys. B {\bf 406} (1993) 19;
H. Fritzsch and Z.Z. Xing, Nucl. Phys. B {\bf 556} (1999) 49.

\end{thebibliography}
\end{document}